\documentclass[conference]{IEEEtran} 

\usepackage[utf8]{inputenc} 
\usepackage[T1]{fontenc}    
\usepackage{hyperref}
\usepackage{booktabs}       
\usepackage{graphicx}
\usepackage{nicefrac}       
\usepackage{microtype}      
\usepackage[cmex10]{amsmath}
\usepackage{amssymb}
\usepackage{amsfonts}
\usepackage{thmtools}
\usepackage{subcaption}
\usepackage[linesnumbered,ruled]{algorithm2e}

\makeatletter
\newcommand{\removelatexerror}{\let\@latex@error\@gobble}
\makeatother

\usepackage{color}

\definecolor{Rafael}{RGB}{139,0,0} 

\definecolor{rick}{RGB}{0,139,0}

\title{Deep Learning for the Gaussian Wiretap Channel}

\author{
	  \authorblockN{Rick Fritschek$^{\ast}$, Rafael F. Schaefer$^{\dagger}$, and Gerhard Wunder$^{\ast}$\\[2mm]}
	  \IEEEauthorblockA{\begin{tabular}{cc}
	       \begin{tabular}{c}
	           $^{\ast}$ Heisenberg Communications and Information Theory Group\\
                        Freie Universit\"at Berlin, \\
                        Takustr. 9,
                        14195 Berlin, Germany\\
                        Email: \texttt{\{rick.fritschek, g.wunder\}@fu-berlin.de}
	       \end{tabular}
	       \begin{tabular}{c}
	           $^{\dagger}$ Information Theory and Applications Chair\\
	                        Technische Universit{\"a}t Berlin \\
                            Einsteinufer 25, 10587 Berlin, Germany\\
                            Email: \texttt{rafael.schaefer@tu-berlin.de}
	       \end{tabular}
	  \end{tabular}
}}

\begin{document}

\maketitle

\begin{abstract}
End-to-end learning of communication systems with neural networks and particularly autoencoders is an emerging research direction which gained popularity in the last year. In this approach, neural networks learn to simultaneously optimize encoding and decoding functions to establish reliable message transmission. In this paper, this line of thinking is extended to communication scenarios in which an eavesdropper must further be kept ignorant about the communication. The secrecy of the transmission is achieved by utilizing a modified secure loss function based on cross-entropy which can be implemented with state-of-the-art machine-learning libraries. This secure loss function approach is applied in a Gaussian wiretap channel setup, for which it is shown that the neural network learns a trade-off between reliable communication and information secrecy by clustering learned constellations. As a result, an eavesdropper with higher noise cannot distinguish between the symbols anymore. 
\end{abstract}

\section{Introduction}
Communication theory addresses the problem of reliably transmitting information and data from one point to another. A communication system itself can be abstracted into three blocks: (1) An encoder at the transmitter side, which takes a message to transmit and encodes it into a codeword. (2) A noisy channel, which transforms the transmitted codeword in a certain way. And (3) a decoder at the receiver side, which estimates the transmitted message based on its noisy channel output. The channel is usually fixed and accounts for example for signal impairments and imperfections in real life scenarios, such as wireless communication. The communication channel of interest in this work is the additive white Gaussian noise channel (AWGN). The aim is now to develop appropriate encoding and decoding schemes to combat the impairments of this channel. This usually involves adding redundancy and introducing sophisticated techniques which use the available communication dimensions (e.g. frequency and time) in an optimal way. However, constructing capacity-achieving coding schemes is a highly non-trivial task even for very simple communication scenarios. It is therefore natural to search for solutions which can simplify this process, for example using deep learning.

Recent developments have shown that a neural network (NN) can simultaneously learn encoding and decoding functions by implementing the communication channel as an autoencoder with a noise layer in the middle, see for example \cite{OShea2017} and references therein. Surprisingly, \cite{SBrink2018} showed that those learned encoding-decoding systems come close to practical baseline techniques. The appeal of this idea is that complex encoding and decoding functions can be learned without extensive communication theoretic analysis. It also enables an on-the-fly ability of the system to cope with new channel scenarios. 

In this paper, we are interested not only in learning the encoding and decoding to account for reliable communication, but also in exploring the possibility to learn how the communication can be secured at the physical layer. To this end, we are interested in physical layer security approaches, which establish a secure transmission by utilizing the intrinsic channel properties and by employing information-theoretic principles as security measures. In particular, these approaches result in secrecy techniques which are independent of the computational power of unknown adversaries, which is in contrast to prevalent computational complexity or cryptographic approaches, e.g. \cite{diffie1976new} and \cite{Rivest:1978}. It therefore results in a stricter notion of security with the drawback that these schemes need to be designed for specific communication scenarios. The simplest scenario which involves reliable transmission and secrecy is the wiretap channel \cite{Wyner75}. This refers to a three-node network with a legitimate transmitter-receiver link and an external eavesdropper which must be kept ignorant of the transmission. It has been shown that specific encoding and decoding techniques exploit an inherent asymmetry (of the additive noise) of the legitimate receiver and the eavesdropper to account for physical layer secure communication. The secrecy capacity of the wiretap channel, i.e., maximum transmission rate at which both reliability and secrecy are satisfied, is known \cite{BlochBarros-2011-PhysicalLayerSecurity,PoorSchaefer-2017-WirelessPhysicalLayerSecurity}. However, constructing suitable encoders and decoders which achieve the secrecy capacity remains a non-trivial challenge. There are several approaches to this problem and most of them are based on polar, LDPC or lattice codes, see \cite{Wu_InfoSecCodingOverview}. However, those techniques are not practical and only work for highly specialized cases/channels. Our motivation is therefore that a NN code design provides a way for on-the-fly code design, which is practical for any channel. The question at hand is now, how to exploit and modify the autoencoder concept to also include the physical layer secrecy measures to obtain coding schemes for physical layer secure communication scenarios. In this paper, we demonstrate that this is indeed possible by creating a training environment where two NN decoders compete against each other. For that we define a novel security loss function, which is based on the cross-entropy. We then show resulting constellations and the probability of error for an SNR range before and after secure coding.

{\bf Related work:}
The work of \cite{OShea2017} introduced the idea of using an autoencoder NN concept to model the communication scenario. The main drawback of this method is, that the channel model needs to be differentiable, which can become a problem with real world channels. However, it was shown in \cite{SBrink2018}, that the learned encoding and decoding rules provide a system which comes close to classical schemes and performs well if used on actual over-the-air transmissions. Moreover, \cite{Hoydis2018} showed, that the training can be done without a mathematical channel model by including reinforcement learning. This shows that end-to-end learning of communication systems can be a viable technology. This also holds for fiber communication as shown in \cite{li2018achievableFiber} and in \cite{karanov2018endFiber}, which utilized the autoencoder concept. Moreover, the concept can be used to learn advanced communication schemes such as orthogonal frequency division multiplexing (OFDM), which enables reliable transmission in multi-path channel scenarios as shown in \cite{felix2018ofdm}. 

The idea of using two competing NNs in a specific context is not new. One of the first works was for example \cite{schmidhuber1992learning} on the principle of predictability minimization. This principle is as follows: for each unit inside a NN exists an adaptive predictor, predicting the unit, based on all other units. The units are now trained to minimize this predictability, therefore enforcing independence between the units. Another popular instance of competing NNs are generative adversarial networks (GANs) as introduced in \cite{goodfellow2014generative}.
There, the two NNs consist of a generative model and a discriminative model, with the later predicting the probability that a sample came from the former model. The generative model is now trained to maximize the error probability of the discriminator model, which introduces an adversarial process. Another recent work is \cite{abadi2016learning}, where a key was provided to Alice and Bob and the NN learns to use the key on its communication link in a way, such that Eve cannot decipher the message (since she has no key). It is therefore a neural cryptography setting and different from our approach, as our network learns to encode a message for direct transmission such that Eve cannot decode it.

{\bf Notation:} We stick to the convention of upper case random variables $X$ and lower case realizations $x$, i.e. $X\sim p(x)$, where $p(x)$ is the probability mass function of $X$. Moreover, we use upper case bold script $\mathbf{X}$ for random vectors and constant matrices, while lower case bold script $\mathbf{x}$ is used for constant vectors. We also use $|\mathcal{X}|$ to denote the cardinality of a set $\mathcal{X}$.

\section{Physical layer security in wiretap channels}
\label{InfoThModel}

The wiretap channel is a three-node network in which a sender (Alice) transmits confidential information to a legitimate receiver (Bob) while keeping an external eavesdropper (Eve) ignorant of it. This setup can be seen as the simplest communication scenario that involves both tasks of reliable transmission and secrecy. Accordingly, this is the crucial building block of secure communication to be understood for more complex communication scenarios.

\begin{figure}
\centering
\includegraphics[scale=0.8]{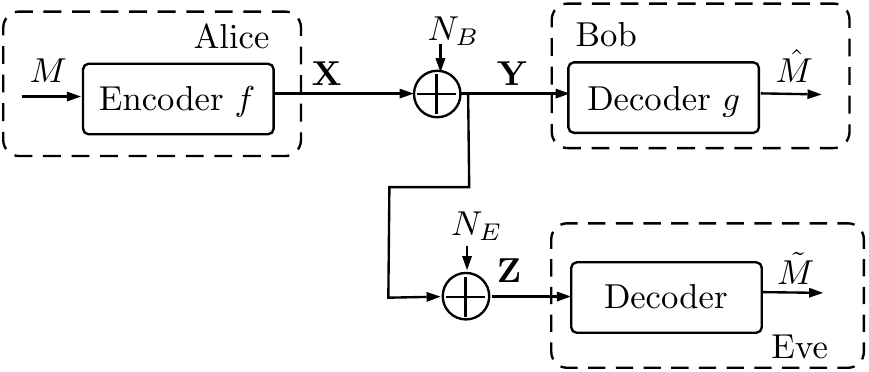}
\caption{Degraded Gaussian wiretap channel. The confidential communication is between Alice and Bob, while Eve tries to eavesdrop upon it.}
\label{Wiretap-degraded-classical}
\end{figure}

In this paper, we study the degraded Gaussian wiretap channel as depicted in Fig. \ref{Wiretap-degraded-classical}. The legitimate channel between Alice and Bob is given by an additive white Gaussian noise (AWGN) channel as
\begin{equation}
	Y_i=X_i+N_{B,i}
    \label{eq:wiretap_y}
\end{equation}
where $Y_i$ is the received channel output at Bob, $X_i$ is the channel input of Alice, and $N_{B,i}\sim \mathcal{N}(0,\sigma_B^2)$ is the additive white Gaussian noise at Bob at time instant $i$. The eavesdropper channel to Eve is then given by
\begin{equation}
	Z_i=Y_i+N_{E,i} = X_i+N_{B,i}+N_{E,i}
    \label{eq:wiretap_z}
\end{equation}
where $Z_i$ is the received channel output at Eve and $N_{E,i}\sim \mathcal{N}(0,\sigma_E^2)$ is the additive white Gaussian noise at Eve. This defines a degraded wiretap channel for which the eavesdropper channel output $Z_i$ is strictly worse than the legitimate channel output $Y_i$.\footnote{Note that any Gaussian wiretap channel of the general form $Y_i=hX_i+N_{B,i}$ and $Z_i=gX_i+N_{E,i}$ with $h$ and $g$ multiplicative channel gains can be transformed into an equivalent degraded wiretap channel as in \eqref{eq:wiretap_y}-\eqref{eq:wiretap_z}. This means that any Gaussian wiretap channel is inherently degraded, cf. for example \cite[Sec. 5.1]{BlochBarros-2011-PhysicalLayerSecurity}.}

The communication task is now as follows: To transmit a message $m\in\mathcal{M}=\{1,...,|\mathcal{M}|\}$, Alice encodes it into a codeword $\mathbf{x}(m)=f(m)$ of block length $n$, where $\mathbf{x}\in \mathcal{X}^n$.\footnote{Usually, $\mathcal{X}=\mathbb{R}$ in the Gaussian setting.} Moreover, we assume an average transmission power constraint $\sum_{i=1}^n x_i^2(m)\leq nP$. At the receiver side, Bob obtains an estimate $\hat{m}$ of the transmitted message by decoding its received channel output as $\hat{m}=g(\mathbf{y})$. The transmission rate is then defined as $R=\log|\mathcal{M}|/n$.

The secrecy of the transmitted message is ensured and measured by information theoretic concepts. There are different criteria of information theoretic secrecy including weak secrecy \cite{Wyner75} and strong secrecy \cite{maurer2000strong}. In the end, all criteria have in common that the output at the eavesdropper $\mathbf{Z}$ should become statistically independent of the transmitted message $M$ implying that no confidential information is leaked to the eavesdropper. For example, strong secrecy is defined as \begin{equation}
\lim_{n \rightarrow \infty} I(M;\mathbf{Z})=0
\label{screcy criterion}
\end{equation}
with $I(M;\mathbf{Z})=\sum_{m,\mathbf{z}}p(m,\mathbf{z})\log\frac{p(m,\mathbf{z})}{p(m)p(\mathbf{z})}$ the mutual information between $M$ and $\mathbf{Z}$, cf. for example \cite{CoverThomas06ElementsInformationTheory}.

The secrecy capacity now characterizes the maximum transmission rate $R$ at which Bob can reliably decode the transmitted message while Eve is kept in the dark, i.e., the secrecy criterion \eqref{screcy criterion} is satisfied while achieving a vanishing error probability $\lim_{n\rightarrow \infty} P_e=\text{Pr}[M \neq\hat{M} ]=0$. The secrecy capacity of the Gaussian wiretap channel is known \cite{BlochBarros-2011-PhysicalLayerSecurity,leung1978gaussian} and is given by
\begin{align}
C_s&=\max_{p(x)}\left(I(X;Y)- I(X;Z) \right)\label{Eq4}\\ 
&=\frac{1}{2}\log\left(1+\frac{P}{\sigma_{B}^2} \right)-\frac{1}{2}\log \left(1+\frac{P}{\sigma_E^2+\sigma_{B}^2} \right).
\end{align}

{\bf Mututal information vs. cross-entropy:}
A straight-forward approach to optimize a NN based on information-theoretic criteria would be an optimization based on the mutual information as in \eqref{screcy criterion} and \eqref{Eq4}. In the security context, this would mean to optimize the encoder $f(\cdot)$ and decoder $g(\cdot)$ mapping to maximize $I(X;Y)$, while minimizing $I(X;Z)$. However, estimating the mutual information from data samples is a non-trivial challenge, due to the unknown underlying distribution. One approach is for example the variational information maximization, introduced in \cite{barber2003algorithm}, which was recently applied to GANs \cite{chen2016infogan}. This technique computes a tractable lower bound to maximize the mutual information between two distributions. However, in our case we would need a technique to simultaneously upper and lower bound two connected mutual information terms. To circumvent this challenge, we adapt a secrecy criterion based on the cross-entropy on which we elaborate further in the next section.

\section{Neural network implementation}
\subsection{General model}

\begin{figure}
\centering
\includegraphics[scale=0.8,angle=90]{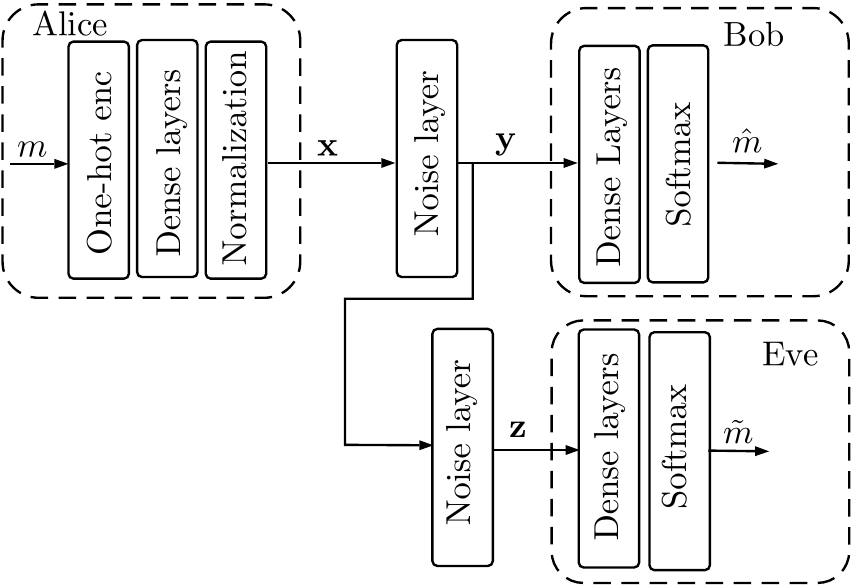}
\caption{Neural network implementation of the degraded wiretap channel.}
\label{WiretapNN}
\end{figure}

As in the reference work \cite{OShea2017}, we implemented the communication scenario using an autoencoder-like network setting. An autoencoder is usually a NN which aims to copy the input $m$ of the network to its output. It consists of an encoder $\mathbf{x}=f(m)$, which maps the input to some codewords and a decoder $\hat{m}=g(\mathbf{x})$ which aims to estimate the input from the output, see \cite{Goodfellow:2016:DL:3086952}. It is therefore a perfect scenario for the communication problem. Usually, these autoencoders are restricted in a certain way, for example that the encoding function performs a dimensionality reduction. That way, the autoencoder can learn useful properties of the data, which are needed for reconstruction. This is in contrast to the communication scenario where the encoder aims to introduce redundancy, i.e. increase the dimensionality. Moreover, there is a noise function in-between encoder and decoder. The NN therefore learns to represent the input in a higher dimensional space to combat the noise corruption, such that the decoder can still estimate the correct output. 

The general structure of our NN can be seen in Fig.~\ref{WiretapNN}. There, the message $m$ gets one-hot encoded into the binary vector $\mathbf{s}\in\mathbb{F}_2^{|\mathcal{M}|}$, which can be viewed as a probability distribution that shows which message was sent and is fed into the NN. The encoder is comprised of two fully connected/dense layers, where the first layer maps $\mathbf{s}$ to $\mathbf{s}'\in\mathbb{R}^{|\mathcal{M}|}$ with a ReLU activation function, while the second layer maps the input to $\mathbf{x}\in \mathbb{R}^n$ with no activation function\footnote{The first layer is given by $\mathbf{s}'=\sigma_{\text{ReLU}}(\mathbf{W}_1\mathbf{s}+\mathbf{b}_1)$ and the second layer by $\mathbf{x}=\mathbf{W}_2\mathbf{s}'+\mathbf{b}_2$ where $\mathbf{W}_i$ and $\mathbf{b}_i$ represent the weight matrices and the bias vectors, respectively and $\sigma_{\text{ReLU}}:=\max(0,\cdot)$.}. Here, $n$ represents the codeword dimension and can be thought of as time instants or channel uses. The last layer of the encoder is a normalization, where the codewords get a unit power constraint, which is either the classical average power constraint over the codeword dimension, i.e.  $\sum_{i=1}^n x_i^2 \leq n$  or a batch average power constraint $\tfrac{1}{N_b}\sum_{j=1}^{N_b}\sum_{i=1}^n x_i^2 \leq n$, where the average is taken over the batch and the codeword dimension. Note that the normalization actually enforces that the resulting codewords have exactly the power specified, turning the inequalities into equalities. For the classical average power constraint, this will yield a constellation, where all the points lie on a circle with radius $\sqrt{n}$. The channel itself is realized as in Section~\ref{InfoThModel}. Here we scale the variance as $\sigma^2=\tfrac{1}{\text{SNR}}$, where SNR means Signal-to-Noise Ratio. Moreover, we have two receiver blocks, which are equally constructed. Each one has two fully connected layers, where the first one maps the channel output back to $\mathbb{R}^{|\mathcal{M}|}$ with a ReLU activation function and the second one maps from $\mathbb{R}^{|\mathcal{M}|}$ to $\mathbb{R}^{|\mathcal{M}|}$ with a linear activation function. As the last step, we use the softmax function\footnote{The softmax function is defined as $f(\mathbf{x}):=\frac{\exp(\mathbf{s}_i)}{\sum_j \exp(\mathbf{s}_j)}$.} and the cross-entropy as loss function. The softmax gives a probability distribution $\hat{\mathbf{s}}\in (0,1)^{|\mathcal{M}|}$ over all messages, which is fed into the cross-entropy:
\begin{IEEEeqnarray}{rCl}
H(p_{\text{data}}(M),p_{\text{model}}(M))&=&-\sum_{m\in \mathcal{M}}p_{\text{data}}(m)\log p_{\text{model}}(m)\IEEEnonumber\\
&=&-\sum_{i=1}^{|\mathcal{M}|} s_i \log \hat{s}_i
\label{cross-entropy-formula}
\end{IEEEeqnarray}

as a loss function, which we then average over the batch sample size $N_b$. This can be seen as a maximum likelihood estimation of the send signal, see for example \cite[Chap.~5]{Goodfellow:2016:DL:3086952}. The index of the element of $\hat{\mathbf{s}}$ with the highest probability will be the decoded symbol $\hat{m}$. The same loss function is applied to the receiver model of the wiretapper, i.e. Eve. However, training for security needs a different loss function.

\subsection{The security loss function}

Remember that information theoretic security results in a difference of mutual information terms between the links. This is in general hard to compute and accordingly difficult use for NN optimization. We therefore focus on the cross-entropy and establish another way to define the security loss. A straightforward method would be to define a security loss function by considering the differences between the cross-entropy losses of Bob and Eve's receiver. The secure loss would therefore be 
\begin{equation}
L:=H(p_{\text{data}}(M),p_{\text{Bob}}(M)) - H(p_{\text{data}}(M),p_{\text{Eve}}(M)).
\end{equation}
Here, $p_{\text{Bob}}$ and $p_{\text{Eve}}$ are the resulting probability mass functions from the softmax output in Bob and Eve's decoder. Note that the data distribution is one-hot encoded. The sum in Eq.~\eqref{cross-entropy-formula} is therefore reduced to $-\log \hat{s}_i$ for a sent symbol $s_i$. Minimizing $H(p_{\text{data}}(M),p_{\text{Bob}}(M))$ trains the encoder to maximize the output probability of symbol $s_i$, and thereby reducing the output probability of all other symbols at Bob. In contrast, maximizing $H(p_{\text{data}}(M),p_{\text{Eve}}(M))$ forces the system to reduce the output probability of the symbol $s_i$ and therefore randomly forces a higher probability on other symbols $\hat{s}_{j\neq i}$, depending on the weights of Eves decoder. This means that training the network for each symbol would result in training the encoder (Alice), such that Eve sees that symbol in another random location, only based on the weight configurations of Eve's network (and therefore highly dependent on the adversary). Moreover, cross-entropy is unbounded from above, resulting in a logarithmic growing negative loss from $H(p_{\text{data}}(M),p_{\text{Eve}}(M))$. This shows that the loss function above is fundamentally ill-suited for the problem. We therefore decided to mimic traditional wiretap-coding techniques. Theoretically, the usual approach is to map a specific message to a bin of codewords and then select a codeword randomly from this bin. There, intuitively, the randomness is used to confuse the eavesdropper. A more concrete method is to use coset codes, where the actual messages label the cosets, and the particular codeword inside the coset is again chosen at random. This method goes back to the work of \cite{Wyner75} and we refer the reader to \cite[Appendix A]{Oggier16} for an introduction. The idea is that Eve can only distinguish between clusters of codewords. Whereas the messages itself are hidden randomly in each cluster. However, the legitimate receiver has a better channel and can also distinguish between codewords inside the clusters. This results in a possible secure communication by trading a part of the communication rate for security. 

Our approach is now that we train the network such that Eve sees the same probability for all codewords in a certain cluster. The security loss function is a sum of the cross-entropy of Bob and the cross-entropy of Eve's received estimated distribution with a modified input distribution. The input distribution is modified in a way, such that clusters of codewords have the same probability. Consider for example the training vector batch $\mathbf{m}=(1,2,3,4)$, resulting in the one-hot encoding matrix
\begin{equation}
\mathbf{S}=
  \begin{bmatrix}
    1 & 0 & 0 & 0 \\
    0 & 1 & 0 & 0 \\
    0 & 0 & 1 & 0 \\
    0 & 0 & 0 & 1 \\
  \end{bmatrix}\end{equation} where the rows are the samples of the batch and the columns indicate the symbol.

\begin{figure}[!t]
 \removelatexerror
{
\begin{algorithm}[H]
    \LinesNumberedHidden
    \SetKwInput{Input}{Input}
    \SetKwInput{Output}{Output}
    \SetKwInput{Initialize}{Initialize}
	\Input{\emph{number of clusters} $l$, \emph{number of symbols} $|\mathcal{M}|$, kmeans.labels, \emph{number of elements in the clusters} $n_c$}
    \Output{\emph{Equalization operator }$\mathbf{E}\in\mathbb{R}^{|\mathcal{M}|\times  |\mathcal{M}|}$}
    \Initialize{$\mathbf{E}=\mathbf{0}$}
        \For{$j\leftarrow 1$ \KwTo $l$}{
       		\For{$i \leftarrow 1$ \KwTo $|\mathcal{M}|$}{
         		\If{kmeans.labels(i)=j}{
         			\For{$k\leftarrow 1$ \KwTo $|\mathcal{M}|$}{
         				\If{kmeans.labels(k)=j}{
         				$\mathbf{E}_{i,k}=\frac{1}{n_c}$
                        }
         			}
                 }
             }
          }
    \caption{This algorithm calculates the operator $\mathbf{E}$, which is used on the input distribution and returns a modified distribution, such that clusters of the input are uniformly distributed.}
    \label{algo:label_equalization}
\end{algorithm}
}
\end{figure}

Multiplying this one-hot input matrix from the right with the matrix $\mathbf{E}$, see Algorithm 1, results in the equalized matrix $\bar{\mathbf{S}}$:
\begin{equation}
 \bar{\mathbf{S}}=\mathbf{S}\mathbf{E}=
  \begin{bmatrix}
    0.5 & 0.5 & 0 & 0 \\
    0.5 & 0.5 & 0 & 0 \\
    0 & 0 & 0.5 & 0.5 \\
    0 & 0 & 0.5 & 0.5\\
  \end{bmatrix}.
  \end{equation}
One can see that in the resulting matrix $\bar{\mathbf{S}}$, the first and second symbol, and the third and fourth symbol have the same distribution. The advantage of this method is, that we only need to calculate an ${|\mathcal{M}|}\times {|\mathcal{M}|}$ matrix, which can be used with any input batch size. The new security loss can therefore be defined as
\begin{IEEEeqnarray}{rCl}
L_{\text{sec}}&:=& (1-\alpha) H(p_{\text{data}}(M),p_{\text{Bob}}(M)) \IEEEnonumber\\
&&+ \alpha H(p_{\overline{\text{data}}}(M)),p_{\text{Eve}}(M))\IEEEnonumber\\
&=&(\alpha-1)\sum_{i=1}^{|\mathcal{M}|} s_i \log \hat{s}_i -\alpha \sum_{i=1}^{|\mathcal{M}|} \bar{s}_i \log \tilde{s}_i,
\label{Loss_Sec}
\end{IEEEeqnarray}

 where $\tilde{\mathbf{s}}$ is Eve's decoded distribution and $\bar{\mathbf{s}}$ stands for the equalized input symbol distribution. Furthermore, the $\alpha$ parameter controls the trade-off between security and communication rate on the legitimate channel. The loss function is then averaged over the batch size $N_b$. Moreover, we chose a modified $k$-means algorithm, which gives equal cluster sizes, for the clustering of the constellation points. This provides us a clustering based on the euclidean distance and goes well with the initial idea of coset codes, when they are implemented as a lattice code with a nearest neighborhood decoder.

\section{Training phases and simulation results}
All of the simulations were done in TensorFlow with the Adam optimizer \cite{kingma2014adam} and gradually decreasing learning rate from $0.1$ to $0.001$. Moreover, we constantly increased the batch size from $25$ to $300$ during the epochs. The training was done with a channel layer of the direct link with an SNR of $12$ dB, and on Eve's link with an SNR of $5$ dB. All of the following figures for constellations and decision regions were done with codeword dimension $n=2$, such that they can be easily plotted in $2$-d. Our training procedure is divided into four phases. 

\begin{figure}
\makebox[\linewidth][c]{%
\begin{subfigure}{.23\textwidth}
  \centering
  \includegraphics[width=\linewidth]{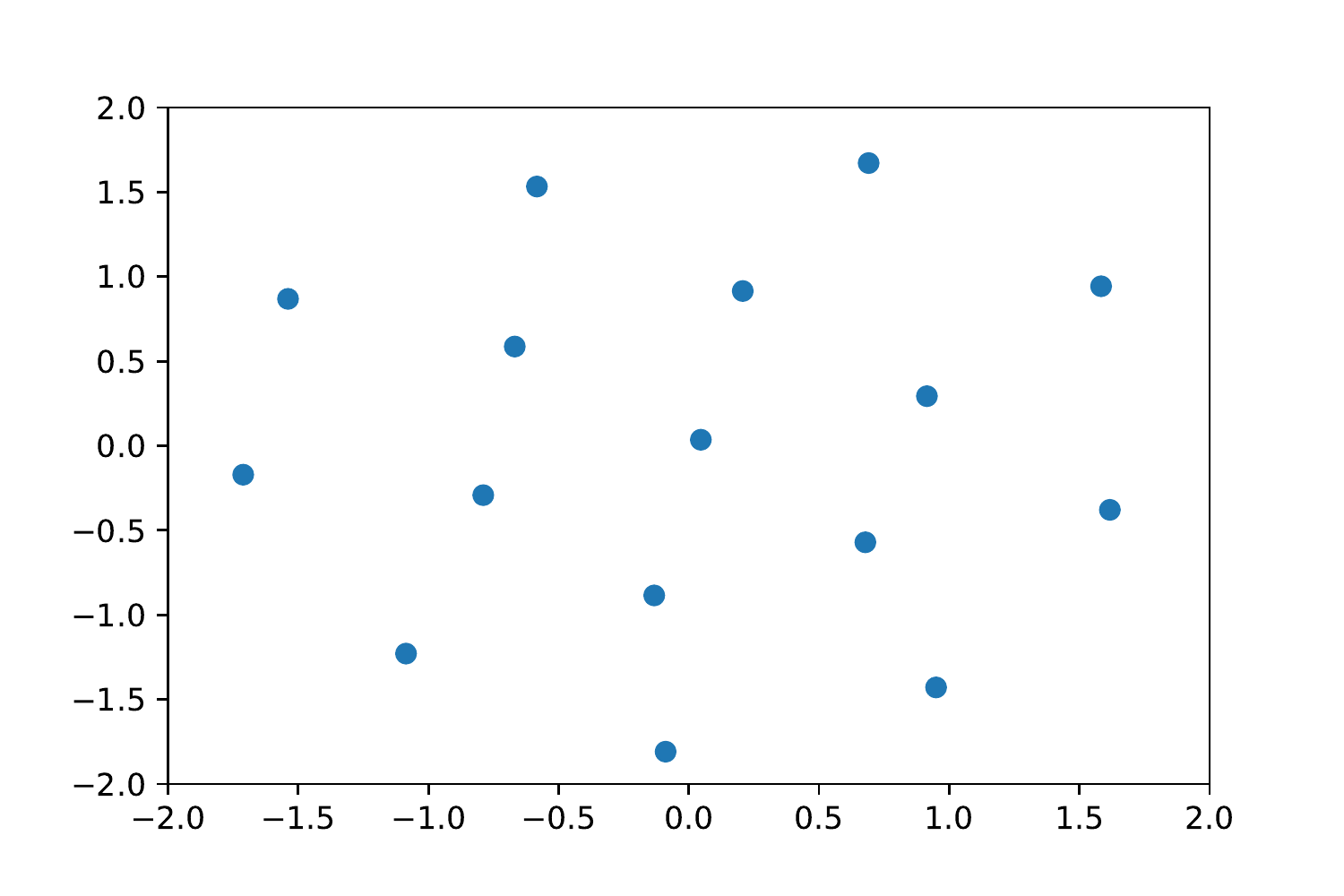}
  \caption{Encoding for batch avg. power constraint.}
\end{subfigure}%
\begin{subfigure}{.23\textwidth}
  \centering
  \includegraphics[width=\linewidth]{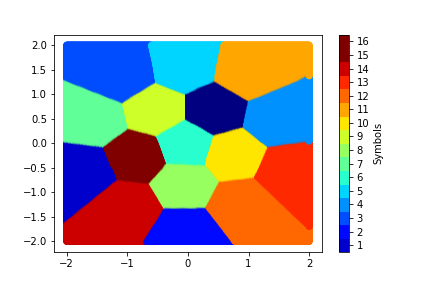}
  \caption{Decoding for batch avg. power constraint.}
\end{subfigure}}
\makebox[\linewidth][c]{%
\begin{subfigure}{.23\textwidth}
  \centering
  \includegraphics[width=\linewidth]{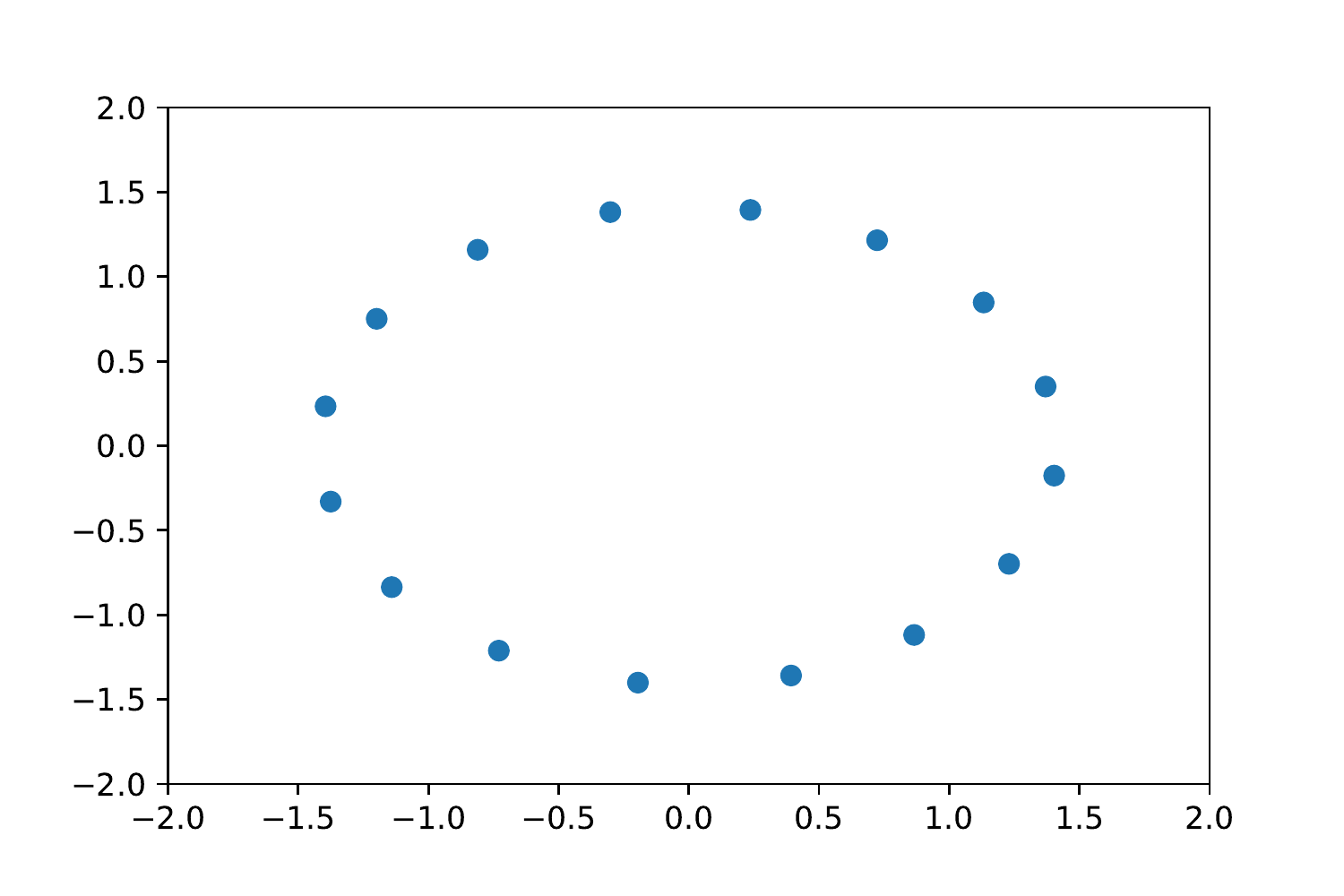}
  \caption{Encoding for avg. power constraint.}
\end{subfigure}%
\begin{subfigure}{.23\textwidth}
  \centering
  \includegraphics[width=\linewidth]{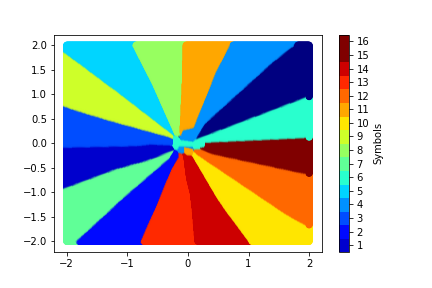}
  \caption{Decoding for avg. power constraint.}
\end{subfigure}}
\caption{The figure shows the learned encoder mappings and decoder decision regions of $16$ symbols for a batch average power constraint and an average power constraint on the symbols.}
\label{fig:Encoding Standard}
\end{figure}

In the first phase, we only train the encoder and the decoder of Bob with the standard cross-entropy, as done in \cite{OShea2017}. Here, the NN learns a signal constellation to combat the noise in an efficient way dependent on the power constraint and the signal-to-noise ratio. The resulting learned constellations, i.e. encoding rules and the learned decision regions, i.e. decoding rules, are shown in Fig.~\ref{fig:Encoding Standard} for an average power constraint over the whole batch and for an average power constraint per symbol.

\begin{figure}
\centering
\begin{subfigure}{0.25\textwidth}
  \centering
  \includegraphics[width=\linewidth]{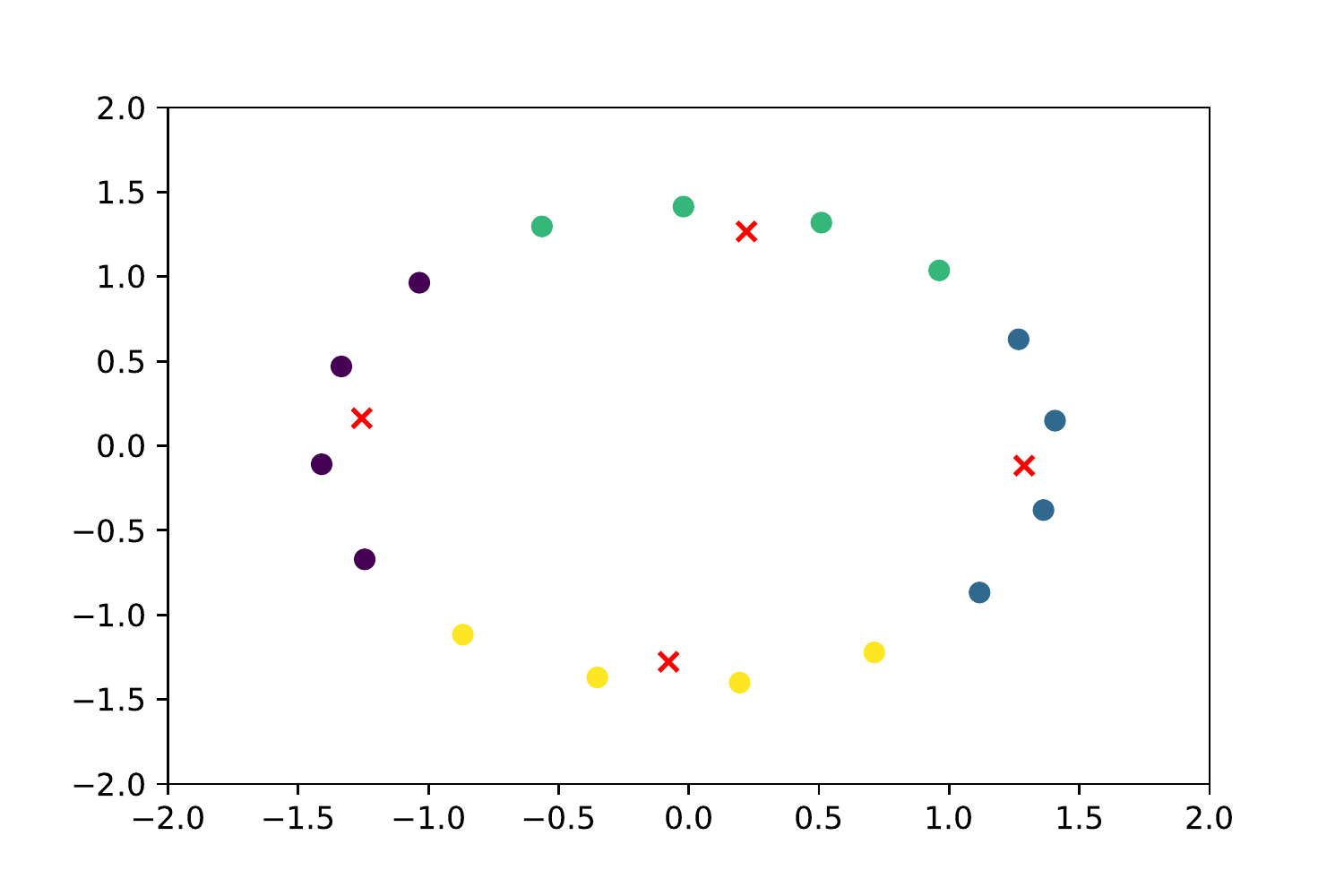}
\end{subfigure}%
\begin{subfigure}{0.25\textwidth}
  \centering
  \includegraphics[width=\linewidth]{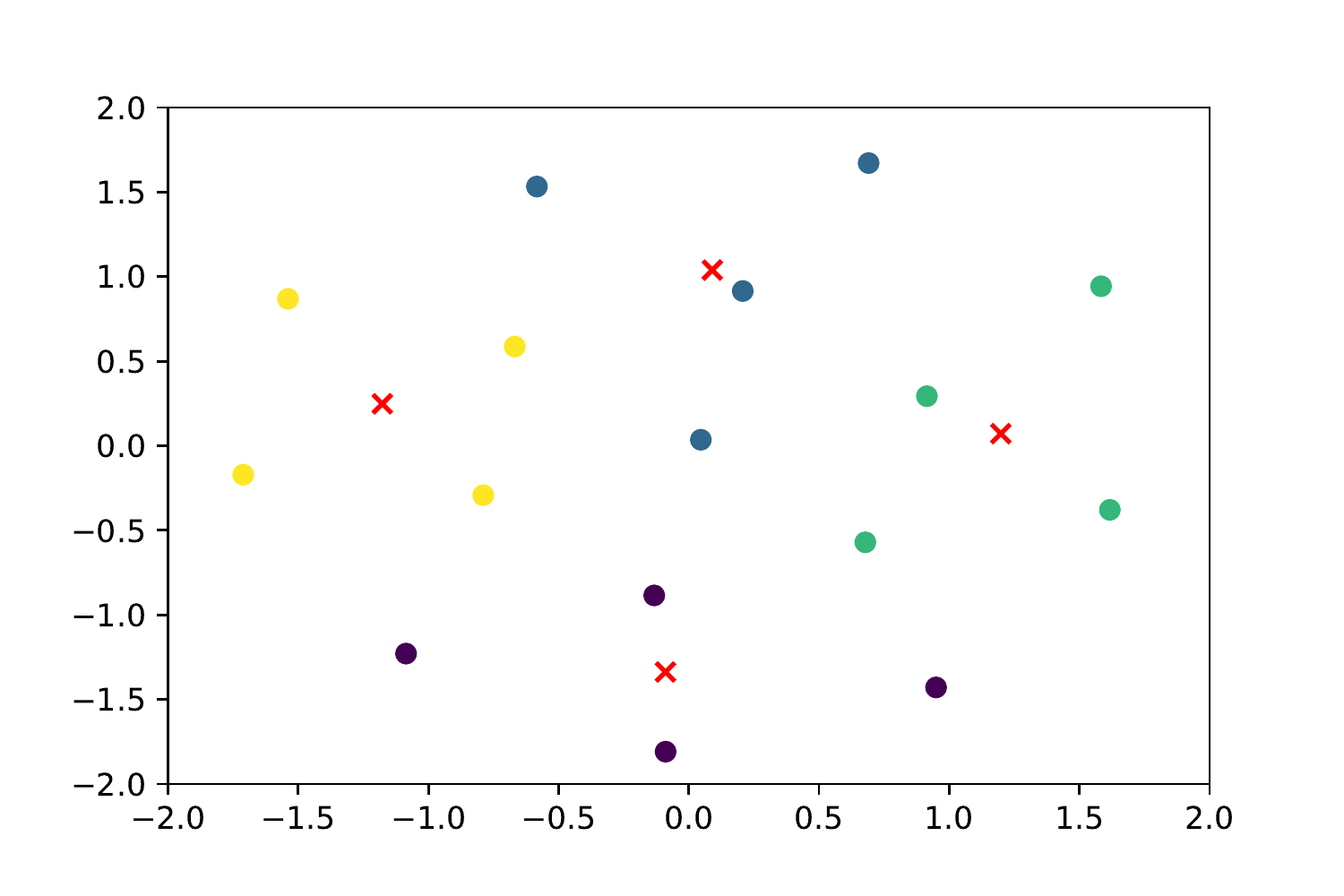}
\end{subfigure}
\caption{The figure shows the clusters for batch average power norm on the right hand side and for average power constraint per symbol on the left hand side. The red crosses show the cluster centers of the $k$-means algorithm.}
\label{fig:Encoding clusters}
\end{figure}
\begin{figure}
\centering
\begin{subfigure}{.25\textwidth}
  \centering
  \includegraphics[width=\linewidth]{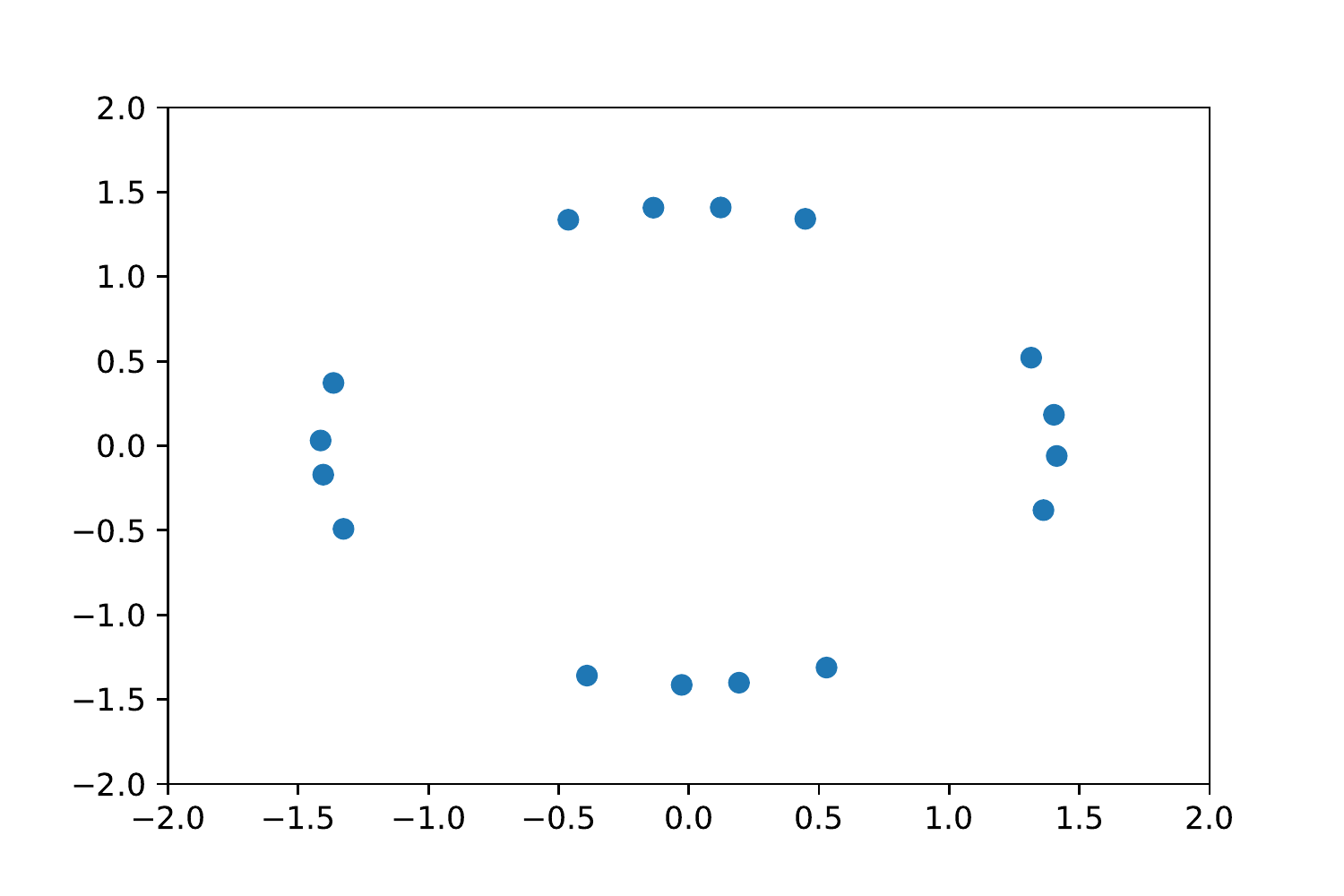}
\end{subfigure}%
\begin{subfigure}{.25\textwidth}
  \centering
  \includegraphics[width=\linewidth]{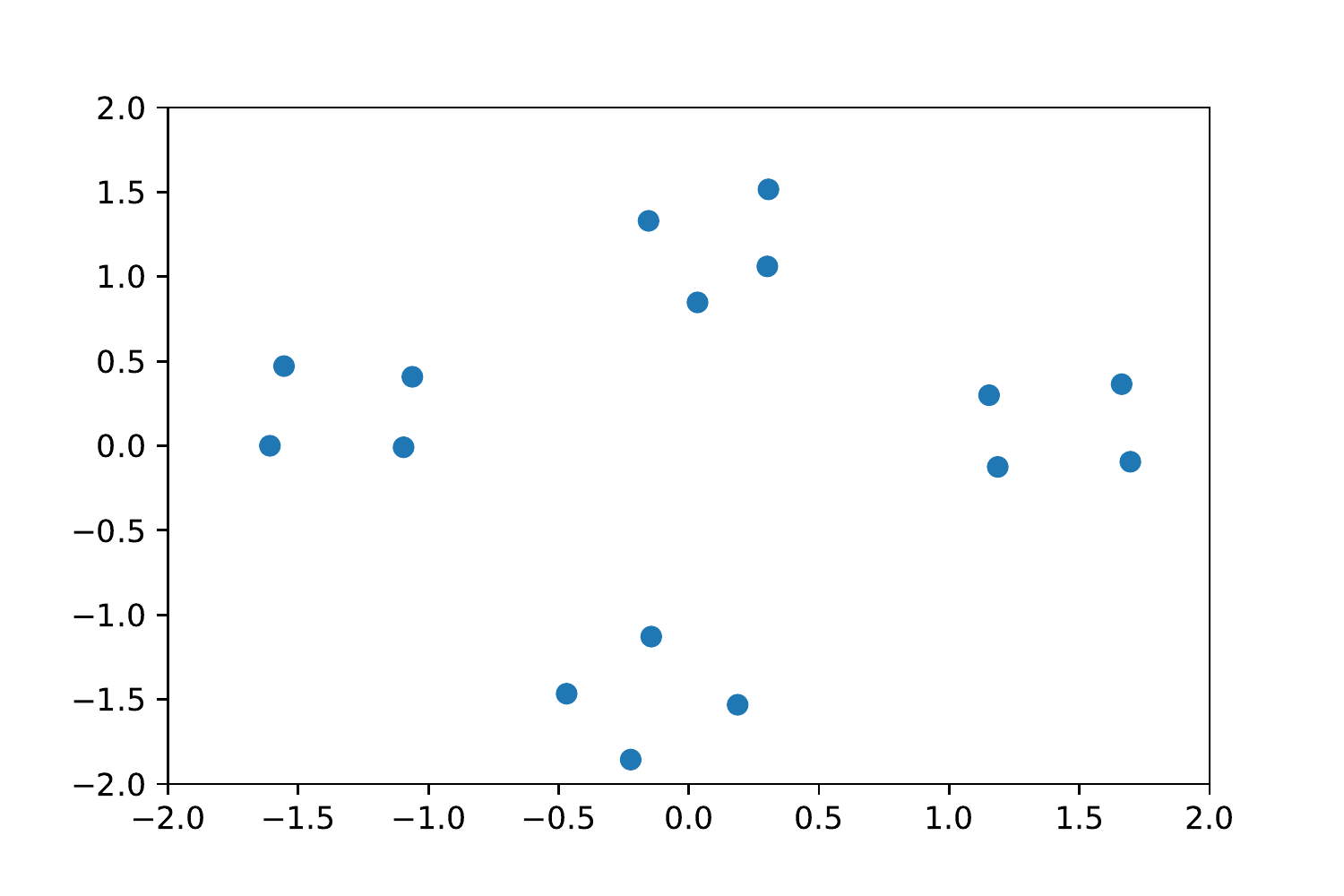}
\end{subfigure}
\caption{The figure shows the learned secure constellations for the decoder with batch average power norm on the right hand side and for average power constraint per symbol on the left hand side. 
}
\label{fig:Encoding sec}
\end{figure}

In the second phase, we freeze the encoding layers, and train Eve on decoding the previously learned encoding scheme with her cross-entropy and the normal input distribution. The reason behind the freezing of the encoder is that we assume that in real-life situations an attacker can not influence the encoding of the signals. We therefore have an unchanged constellation, but Eve's NN learns a decision region for her channel model, i.e. an additional noise factor with $5$ dB SNR.

In the third phase, we freeze the decoding layers of Eve and train the NN with the loss function \eqref{Loss_Sec} and $\alpha=0.7$. This time the freezing is done because we cannot assume that a communication link has access to the decoding function of an attacker. For the equalization of the input clusters, we feed the network with an identity matrix $I_{|\mathcal{M}|}$ and calculate the clusters of the constellation points with the $k$-means algorithm. Fig.~\ref{fig:Encoding clusters} shows the results of the clustering algorithm on the learned constellations. We then use Alg.~\ref{algo:label_equalization} to calculate $\mathbf{E}$ and create the equalized batch label matrix with $\bar{\mathbf{S}}=\mathbf{S}\mathbf{E}$. The resulting training effect is, that the NN tries to pull codewords from the same cluster together, close enough that Eve cannot distinguish the symbols in a cluster, and loose enough such that Bob can still decode them. Fig.~\ref{fig:Encoding sec} shows the learned secure encoding schemes.

After the secure training phase we train Bob's and Eve's decoder once again. Therefore, both neural networks are trained to decode the new secure encoded signals. Hence, we also want to validate that we have indeed achieved security in comparison to a training advantage where Eve was merely not trained enough to decode the new secure encoded symbols.

\begin{figure}
\centering
\includegraphics[scale=0.65]{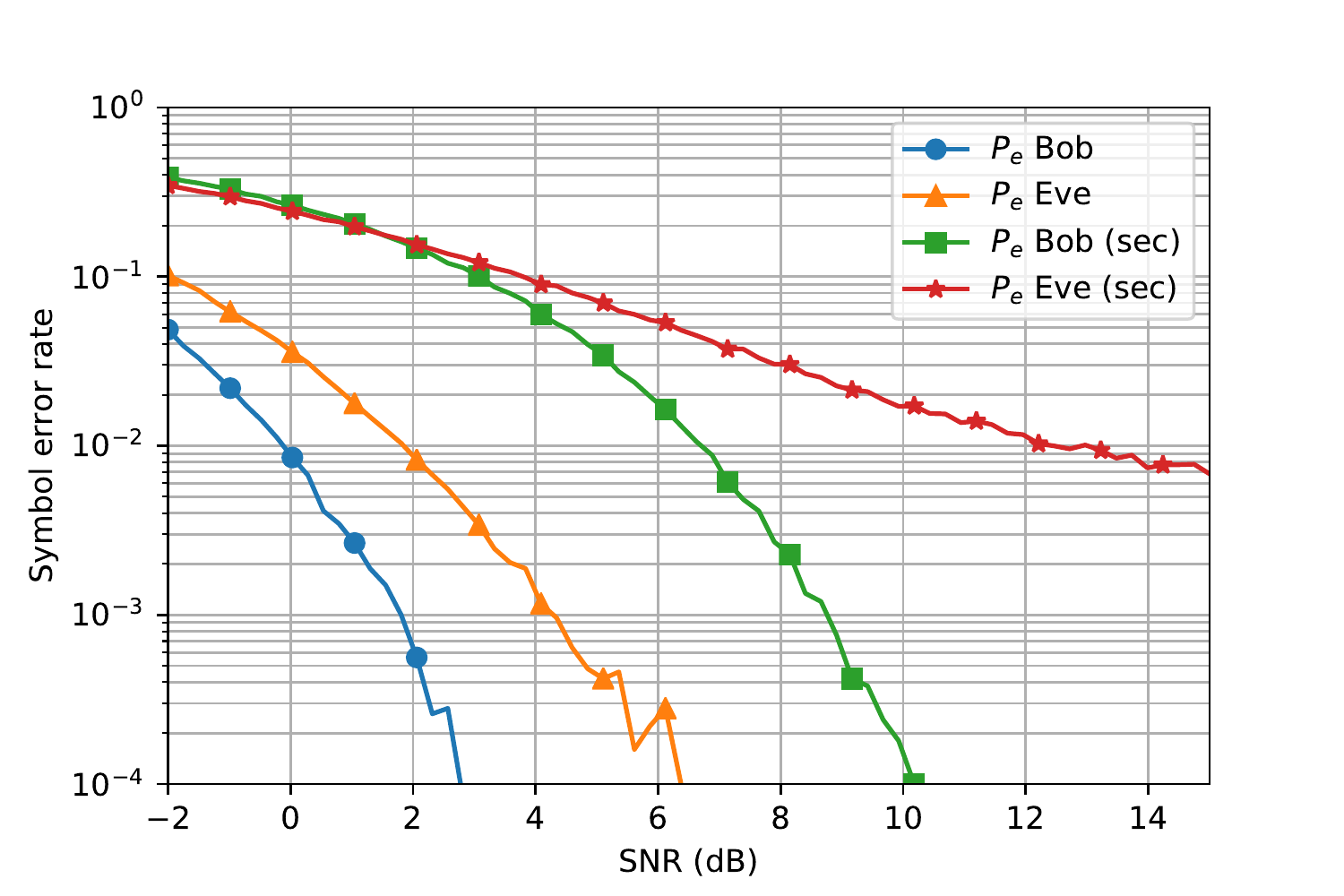}
\caption{The figure shows the symbol error rate to SNR graph for a $16$ symbol constellation size with $n=32$ channel uses and fixed additional SNR of $17$ dB in Eve's channel, before and after secure encoding. Note that the SNR is per symbol and not per bit ($E_B/N_0$).}
\label{Rate_Sec}
\end{figure}

The training phase for $16$ symbols gets more accurate and faster with increasing codeword dimensions $n$, which suggests that the NN can find better constellations. The drawback is that the actual communication rate drops, since the system needs more time instants to transmit a bit. Therefore, we have taken a conservative approach, which resulted in $n=32$. Moreover, we implemented a coset coding algorithm. In our case, we use $4$ clusters, so each cluster consists of $4$ symbols. We have $4$ secure messages, and for every message, a specific cluster is chosen at random (i.e. point in its coset), mapping the $4$ messages to $16$ symbols. This randomness increases Eve's confusion about the transmission. Intuitively, we sacrifice a part of the transmission rate, i.e we only use 4 instead of 16 symbols, to accommodate randomness to confuse Eve. Again we refer to \cite[Appendix A]{Oggier16}, which discusses Eve's error probability and shows that coset coding does increase confusion at the eavesdropper. The NN therefore learns a constellation which can be seen as a finite lattice-like structure, on which one can implement the idea of coset coding. For the actual simulation, we have taken a direct SNR of $10$ dB and an additional SNR of $7$ dB in the adversary link, during the training phase. We then evaluated the symbol error rate which approximates $P_e$ for decoding the symbols before the third training phase, i.e. before secure coding and after all the training, which results in Fig.~\ref{Rate_Sec}. We note that the figures were tested with the same total sample size of $50000$, while the testing was also done with a SNR of $7$ dB in the adversary link. Moreover, we used a factor of $\alpha=0.3$ in the security loss function. 
One can see in the results, that the NN learns a trade-off between communication rate and security.

\section{Conclusions and outlook}
We have shown that the recently developed end-to-end learning of communication systems via autoencoder neural networks can be extended towards secure communication scenarios. For that we introduced a modified loss function which results in a trade-off between security and communication rate. In particular, we have shown that the neural network learns a clustering, which resembles a finite constellation / lattice, which can be used for coset encoding as demonstrated. This opens up the ongoing research of end-to-end learning of communication systems to the field of secure communication, as classical secure coding schemes can be learned and applied with a neural network. We think that our approach via the new loss function, is a fruitful direction in that regard. However, an optimal way would be to tackle the problem by direct optimization via mutual information terms which remains an open problem.

\medskip

\small

\bibliographystyle{./IEEEtran}
\bibliography{./ref}

\end{document}